\documentclass[DIV=16,twocolumn]{scrartcl}
\usepackage[utf8]{inputenc}
\usepackage{cite}
\usepackage{booktabs}

\usepackage{graphicx}
\graphicspath{{../figs/}}
\usepackage{tikz,xspace}

\usepackage{amsfonts}
\usepackage[cmex10]{amsmath}
\usepackage{array}
\usepackage{url}
\usepackage{amssymb}
\usepackage{flushend}

\newcommand{\req}[1]{Eq.\,(\ref{#1})}

\renewcommand{\vec}[1]{\underline{#1}}

\makeatletter
\newsavebox{\@euflag}
\sbox{\@euflag}{\raisebox{-9mm}{\resizebox{!}{1.2cm}{\begin{tikzpicture}
\fill[fill={rgb,255:red,0; green,51; blue,153}] (-27, -18) rectangle (27,18);  
\pgfmathsetmacro\inr{tan(36)/cos(18)}
\foreach \i in {0,1,...,11} {
  \begin{scope}[shift={(30*\i:12)}]
    \fill[fill={rgb,255:red,255;green,204;blue,0}] (90:2)
	\foreach \x in {0,1,...,4} { -- (90+72*\x:2) -- (126+72*\x:\inr) };
  \end{scope}
}
\end{tikzpicture}
}}}
\newcommand\euflag{\usebox{\@euflag}\xspace}
\makeatother

\begin{document}

\title{%
  QBM -- Mapping User-Specified Functions\\
  to Programmable Logic\\
  through a QBF Satisfiability Problem
}

\author{
  Thomas B. Preußer\\
  Marie-Skłodowska-Curie-Fellow\footnote{%
    \euflag\parbox[t]{.72\linewidth}{%
      This project has received funding from the European Union's Framework
      Programme for Research and Innovation Horizon 2020 (2014-2020) under
      the Marie Skłodowska-Curie Grant Agreement No.\,751339.}
  }, Xilinx Ireland\\
  \textsl{thomas.preusser@utexas.edu}
}
\date{June 25, 2017}

\maketitle

\begin{abstract}
  This is a brief overview on the background behind the test set
  formulas generated by the QBM tool. After establishing its
  application context, its formal approach to the generation of QBF
  formulas and the concrete test set formulas are described. Finally,
  some related work will be credited and the source to obtain the
  open-source tool will be identified.
\end{abstract}

\section{Application Context}\label{secIntro}
Programmable logic devices enable users to exploit the tremendous
concurrency benefits of hardware in their compute solutions without
requiring them to go through a silicon production process. The most
powerful among these devices are field-programmable gate arrays
(FPGAs) that provides hundreds of thousands of logic slices, each one
of the capable of computing any arbitrary boolean 6-input function
\cite{ug474}. The configuration of these devices is
computed in a synthesis process, which translates a behavioral
or functional specification into a structural netlist consisting of
the configurable primitives found on the devices and their
configurations. These configurations include, for instance, truth
tables to be stored in lookup table (LUT) components as well as
datapath configurations that establish the proper signal connections.

The synthesis process may start from traditional hardware description
languages like VHDL \cite{vhdl}, from high-level imperative languages
like C++ \cite{vivado_hls} or from functional description in
languages like Chisel \cite{bachrach:2012}. In any case, the specified
behavior will result
in combinatorial equations that in their entirety define the function
for the transition from one system state to the subsequent one also
considering external inputs and computing outputs as specified. The
efficient mapping of these combinatorial functions to the structural
capabilities of the targeted device architecture is crucial as it
defines the costs of hardware and power, and even limits the
application range of a device. Heuristics transforming data structures
like inverter graphs (AIG) \cite{mishchenko:2008} are used to balance or
reshape the dataflow of the combinatorial computation so that it fits
well onto the target structure. Decomposition approaches may be used
to identify the utility of special-purpose resouces such as carry
chains for the implementation of a given function \cite{preusserb:2010}.

The mapping heuristics used by modern synthesis tools
are very mature and deliver high-quality results. However, there is a
growing number of data center applications, most prominently neural
network inference \cite{umuroglu:2017}, which rely on a highly parallel
computation carried out by massively replicated operator cores. For
them, it is a paramount benefit to use the optimal rather than just a
very good implementation of this operator. This makes it also
worthwhile to invest extra effort that goes significantly beyond what
would be acceptable in a regular synthesis run.

\section{QBF Formulation}
The mapping of a user-specified function to a given reconfigurable
logic structure can be expressed as a QBF satisfiability problem as
detailed below. Solving this problem yields precise statements:
\begin{description}
\item[UNSAT]
  that there is no implementation within the range of the granted resources, or
\item[SAT]
  that such an implementation, indeed, exists.
\end{description}
It is remarkable that a satisfying assignment of the variables under the
outermost existential quantifier of the formulas -- which all follow an
EAE pattern -- corresponds directly to an implementing configuration of
the programmable circuit. On the other hand, a problem that has been
found unsatisfiable is no longer worth pondering as no engineering
effort and experience will yield a solution under the given resource
constraints.

A digital combinational reconfigurable circuit can be modeled by its
characteristic function $\Phi$ mapping its $k$ boolean configuration
variables and its $l$ boolean inputs to its $m$ boolean outputs:\vspace*{-1ex}
\begin{align}\label{eqCharacteristic}
  &\Phi:& B^k\times B^l &\to B^m\\\nonumber
  &  & (\vec{c}, \vec{x}) &\mapsto \vec{y}\\[.5ex]\nonumber
  &  &\mbox{with\quad}&\vec{c}=c_0, \dots, c_{k-1}\\\nonumber
  &               &&\vec{x}=x_0, \dots, x_{l-1}\\\nonumber
  &               &&\vec{y}=y_0, \dots, y_{m-1}\\\nonumber
  &               &&B=\{0,1\}
\end{align}
The arguments \vec{c} specify a concrete \emph{configuration} of the
circuit. They are provided through internal SRAM cells or fuses. Their
update is relatively slow or impossible altogether. Hence, the
configured values are kept constant over much or even all the
operation time.

The arguments \vec{x}, on the other hand, are typical functional
inputs that are provided dynamically in the course of operation. They
are evaluated in the context of the current configuration, which
defines the effective functionality of the circuit. The circuit is
capable of implementing a target user function
\mbox{$f: \vec{x}\mapsto\vec{y}$} if and only if:
\begin{equation}\label{eqImplConfig}
  \exists\vec{c}.\;\forall\vec{x}.\;\Phi(\vec{c}, \vec{x})=f(\vec{x})
\end{equation}

A \emph{satisfying assignment} \mbox{$C: \vec{c}\mapsto B^k$}
identifies an \emph{implementing configuration} for $f$.

\req{eqImplConfig} is already a valid formulation of the problem
of mapping a target user function $f$ to a configurable circuit with
the characteristic function $\Phi$. Both functions can be joined into
a single formula:
\begin{equation}
  \exists\vec{c}.\;\forall\vec{x}.\;F'(\vec{c}, \vec{x})
\end{equation}
A very practical format of $F'$ is a conjunction of clauses. This
makes it a collection of constraints that must be all satisfied. It
can, thus, be constructed incrementally. This property is
very valuable as it enables the compilation of the specified model
into its QBF formulation processing statement after statement.

So as to further simplify the formulation of $F'$, it is helpful to
introduce internal node variables $\vec{n}$ that allow the composition
of a constraint from simpler subterms or
subcomponents. Such node variable may, for instance, represent the
interconnections to a component in a structural design. So, the final
structure of our problem statement is:
\begin{equation}\label{eqEAESpec}
  \exists\vec{c}.\;\forall\vec{x}.\;\exists\vec{n}.\;F(\vec{c},\vec{x},\vec{n})
\end{equation}
Solving \req{eqEAESpec} then determines whether or
not there is a configuration of $\vec{c}$ of the configurable circuit
so that for all input combinations over $\vec{x}$, there is an
assignment to the internal node variables $\vec{n}$ so that:
\begin{itemize}
\item
  the circuit works physically correct and satisfies all component
  specifications, and
\item
  the circuit implements the specified user function $f$.
\end{itemize}
Note that $F$ contains two parallel models of how the inputs $\vec{x}$
relate to the outputs $\vec{y}\subset\vec{n}$. One model
specifies the relation as defined by the user and the other relates the
same inputs to the outputs via the configurable circuit
description. Observe that the outputs $\vec{y}$ are subsumed by
the internal node variables in this formulation. Also note that it is
only their role that differentiates the variables into the three sets
$\vec{c}$, $\vec{x}$ and $\vec{n}$. Otherwise, they may all be
viewed as signals or wires carrying a single bit.

For producing a problem statement in QDIMACS format, $F$ must be
specified in conjunctive normal form. This normal form is produced
internally by the QBM tool, which is freely available on
GitHub\footnote{\url{https://github.com/preusser/qbm}}.
The engineer can specify both the configurable circuit as
well as the desired function by structural and behavioral descriptions
as described by Preußer and Erxleben \cite{preussera:2016}.

\section{Test Set Formulas}
The test set formulas are problem statements that ask for the
configuration of the lookup tables and for the selection of their
appropriate inputs so as to implement binary word adders of different
bitwidth within a Xilinx carry-chain structure \cite{ug474}. All problem
statements are satisfiable. The test set includes adder width of four
through seven bits. Recent experiments have shown that this range can
be expected to span from rather simple all the way to currently
infeasible problem complexities \cite{preussera:2016}.

The test set includes, for each problem size, three different
formulations for the selection of the inputs to a lookup table from
the overall set of inputs, which includes logic one and zero in
addition to the actual operand words:
\begin{enumerate}
\item
  Each LUT input is driven by a configurable multiplexer (CMUX)
  whose configuration determines, which one of all the available input
  signals is passed through.
\item
  The first LUT input is driven by a CMUX selecting one from all the
  available input signals. The CMUX modeled for each subsequent LUT input
  can only choose from one fewer input signal than the predecessor.
\item
  A \verb+CHOOSE+ operator is driving all the inputs of the LUT.
  Its configuration determines, which $k$ of all the available inputs
  are forwarded. For each possible selection, it only allows one single
  permutation.
\end{enumerate}
The first option is a naive approach. It models the device capabilities
well but introduces practically useless degrees of freedom. For instance,
it is irrelevant in what order inputs are provided to a LUT, and it is
pointless to feed the same input twice. The second option takes a simple
approach to reduce the available degrees of freedom somewhat but only to
a degree that each choice for a LUT input can be made individually. The
third and last option goes the furthest. However, it is based on identifying
the one out of $\left(\begin{array}{c}n\\k\end{array}\right)$ combinations
that selects the needed set of $k$ inputs from the $n$ available ones.

So as to appreciate the differences of these different models of the input
selection, consider the number of configuration variables that is needed to
encode this selection. In the first naive model, we require
$k\cdot\left\lceil\log n\right\rceil$ input variables; in the third case,
$\left\lceil\log\left(\begin{array}{c}n\\k\end{array}\right)\right\rceil$
ones. We have:
\begin{align}
k\cdot\left\lceil\log n\right\rceil
	&\ge\left\lceil k\cdot\log n\right\rceil\\
	&=\left\lceil\log n^k\right\rceil\\
	&\ge\left\lceil\log\left(\begin{array}{c}n\\k\end{array}\right)\right\rceil
\end{align}
Observe that the final inequality captures a gap that grows significantly
with $k$.

The drawback of the third approach is that the encoding of the
configuration for the selection spans all configuration variables and
becomes part of all the clauses that must be generated. The local
configurations of the first two approaches, on the other hand, comprise
at most $\left\lceil\log n\right\rceil$ variables producing much shorter
clauses.

\section{Related Work}
Previous formal approaches to the boolean matching of combinational
computations to FPGA structures were based on a SAT expansion of the
problem \cite{ling:2007,safarpour:2006}. These approaches are, thus,
able to benefit from the enormous advances in the technology of SAT
solving \cite{jarvisalo:2012}. However, the formulation of the
matching problem in SAT also implies a heavily inflated problem
specification.

This paper demonstrates how the mapping problem is more naturally
formulated as a QBF. This approach avoids the inflation of the problem
specification by literally reiterating the set of the defining clauses
with fresh variables for every possible input combination. Note that
this inflation is, in fact, exponential in the number of Boolean
inputs. Internally, QBF solvers are not unlikely to
utilize SAT solvers themselves \cite{SATinQBF,pigorsch:2010}.
However, exposing the natural structure of the problem to the solver
and using a solver specialized for this structure, can only benefit
the developed tool. The solvers are free in the organization of their
search for a solution and may cut away parts of it on the basis of
their domain-specific knowledge. An impetuous naive SAT expansion can
only disguise higher-level information on the problem structure and
neglects the existence of appropriate tools.

\section{Availability}
QBM is open-sourced and available at
\url{https://github.com/preusser/qbm}. It is published together with
quick starting instructions and simple example problems like the
presented adders.

\bibliographystyle{IEEEtran}
\bibliography{IEEEabrv,paper}

\end{document}